\documentclass[twoside,aps,showpacs,floatfix]{revtex4}

\usepackage{amsmath,amssymb,graphicx,dsfont}

\newcommand{\al}{\alpha}
\newcommand{\vs}{\vec{S}}
\newcommand{\vsi}{\vec{\sigma}}
\newcommand{\rh}{\hat{r}}

\begin{document}

\title{Replica mean-field theory for Levy spin-glasses }

\author{A. Engel}

\email{engel@theorie.physik.uni-oldenburg.de} 
\affiliation{Institut f\"ur Physik, Carl-von-Ossietzky-Universtit\"at,
     26111 Oldenburg, Germany }


\begin{abstract}
Infinite-range spin-glass models with Levy-distributed interactions
show a freezing transition similar to disordered spin systems on
finite connectivity random graphs. It is shown that despite 
diverging moments of the local field distribution this transition can
be analyzed within the replica approach by working at imaginary
temperature and using a variant of the replica method developed for
diluted systems and optimization problems. The replica-symmetric 
self-consistent equation for the distribution of local fields
illustrates how the long tail in the distribution of coupling
strengths gives rise to a significant fraction of strong bonds per
spin which form a percolating backbone at the transition temperature. 
\end{abstract}

\maketitle

\section{Introduction}
Spin-glasses are model systems of statistical mechanics in which
simple degrees of freedom interact via random couplings drawn from a
given probability distribution \cite{BiYo}. The ensuing interplay 
between disorder and frustration gives rise to peculiar static and
dynamic properties which made spin-glasses paradigms for complex
systems with competing interactions. In this way the concepts and techniques
developed for their theoretical understanding \cite{MPV} became
invaluable tools in the quantitative analysis of problems originating
from such diverse fields as algorithmic complexity
\cite{RemiNat,MPZ,HaWe}, game theory \cite{DiOp,EnBe},
artificial neural networks \cite{EnvB}, and cryptography \cite{KaKi}. 

In the present note we study a spin-glass for which the 
couplings strengths are drawn from a Levy-distribution. The main
characteristic of these distributions are power-law tails resulting in 
diverging moments. Compared to the extensively studied spin-glass
models with Gaussian \cite{SK} or other finite moment distributions
\cite{ViBr,KaSo,MePa} Levy-distributed couplings are interesting
for several reasons. On the one hand the comparatively large fraction
of strong bonds gives rise to a mechanism for the glass transition which is
different from the usual scenario. On the other hand these systems
pose new challenges to the theoretical analysis because the diverging
second moments invalidate the central limit theorem which is at the
bottom of many mean-field techniques. Related issues of interest include
the spectral theory of random matrices with Levy-distributed entries
\cite{ranmat1,ranmat2} 
and relaxation and transport on scale-free networks \cite{AlBa}. It is
also interesting to note that the characteristic properties of the
Cauchy-distribution have recently enabled progress in the
mathematically rigorous analysis of matrix games with random pay-off 
matrices \cite{Roberts}.  

The model considered below with the help of the replica method was 
analyzed previously by Cizeau and Bouchaud (CB) using the cavity
approach \cite{CiBo}. In their paper CB remark that they resorted to
the cavity method because they were not able to make progress within
the replica framework. This might have been caused by the fact that at
that time the
central quantity in the replica method was the second moment
of the local field distribution, the so-called Edwards-Anderson
parameter \cite{EdAn}, which for Levy-distributed couplings is likely
to diverge. Later a variant of the replica method 
was developed to deal with non-Gaussian local field distributions
characteristic for diluted spin glasses and complex optimization
problems \cite{Remi}. Until now this approach was 
used only in situations where the local field distribution is
inadequately characterized by its second moment alone and higher
moments of the distribution are needed for a complete description. 
Here we show that the method may also be adapted to situations where
the moments not even exist. 

\section{The model}

We consider a system of $N$ Ising spins $S_i=\pm 1,\, i=1,...,N$ with
Hamiltonian 
\begin{equation}
  \label{eq:H}
  H(\{S_i\})=-\frac1 {2N^{1/\al}} \sum_{(i,j)} J_{ij} S_i S_j\; ,
\end{equation}
where the sum is over all pairs of spins. The couplings $J_{ij}=J_{ji}$
are i.i.d. random variables drawn from a Levy distribution $P_\al(J)$
defined by its characteristic function \cite{Levydist}
\begin{equation}
  \label{eq:defP}
  \tilde{P}_\al(k):=\int dJ \; e^{-ikJ}\; P_\al(J)=e^{-|k|^\al} \; 
\end{equation}
with the real parameter $\al\in (0,2]$. The thermodynamic properties
of the system are described by the ensemble averaged free energy
\begin{equation}
  \label{eq:deff}
  f(\beta):=-\lim_{N\to\infty}\frac{1}{\beta N}\overline{\ln
    Z(\beta)}\; , 
\end{equation}
with the partition function 
\begin{equation}
  \label{eq:defZ}
  Z(\beta):=\sum_{\{S_i\}}\exp(-\beta H(\{S_i\}))\; .
\end{equation}
Here $\beta$ denotes the inverse temperature and the overbar stands for the
average over the random couplings $J_{ij}$.

\section{Replica theory}
To calculate the average in (\ref{eq:deff}) we employ the replica
trick \cite{EdAn} 
\begin{equation}
  \label{eq:replicatrick}
  \overline{\ln Z}=\lim_{n\to 0}\frac{\overline{Z^n}-1}{n}\; .
\end{equation}
As usual we aim at calculating $\overline{Z^n}$ for integer $n$
by replicating the system $n$ times, $\{S_i\}\mapsto \{S^a_i\},\,
a=1,...,n$, and then try to continue the results to real $n$ in order
to perform the limit $n\to 0$.  

Due to the algebraic decay $P_\al(J)\sim |J|^{-\al-1}$ of the
distribution $P_\al(J)$ for large $|J|$ the average
$\overline{Z^n(\beta)}$ does not exist for real $\beta$. On the other
hand, for a purely imaginary temperature,
$\beta=-ik,\,k\in\mathds{R}$,  we find from the very definition of
$P_\al(J)$, cf. (\ref{eq:defP}) 
\begin{equation}
  \label{eq:Zn}
  \overline{Z^n(-ik)}=\sum_{\{S_i^a\}} \exp
\Big(-\frac{|k|^\al}{2N}\sum_{i,j} \Big |\sum_a S_i^a S_j^a\Big |^\al 
     + {\cal{O}}(1)\Big) \; .
\end{equation}
Note that the replica Hamiltonian is extensive which justifies
a-posteriori the 
scaling of the interaction strengths with $N$ used in (\ref{eq:H}). 
The determination of $\overline{Z^n}$ can be reduced to an effective
single site problem by introducing the distributions  
\begin{equation}
  \label{eq:c}
  c(\vs)=\frac1 N \sum_i\delta(\vs_i,\vs)\; ,
\end{equation}
where $\vs=\{S^a\}$ stands for a spin vector with $n$ components. We 
find after standard manipulations \cite{Remi} 
\begin{equation}
  \overline{Z^n(-ik)}=
  \int\prod_{\vs}dc(\vs)\delta(\sum_{\vs}c(\vs)-1) 
  \exp\Big(-N\Big[\sum_{\vs}c(\vs)\ln c(\vs)+\frac{|k|^\al}{2}
  \sum_{\vs,\vs'}c(\vs)c(\vs') |\vs\cdot\vs'|^\al\Big]\Big)\; .\label{eq:Zn2}
\end{equation}
In the thermodynamic limit, $N\to\infty$, the integral in
(\ref{eq:Zn2}) can be calculated by the saddle-point method. The
corresponding self-consistent equation for $c(\vsi)$ has the form
\begin{equation}
  \label{eq:saddle}
  c(\vsi)=\Lambda(n)\exp\Big(-|k|^\al
     \sum_{\vs}c(\vs)|\vs\cdot\vsi|^\al\Big)  \; ,
\end{equation}
where the Lagrange parameter $\Lambda(n)$ enforces the constraint 
$\sum_{\vs} c(\vs)=1$ resulting from (\ref{eq:c}). 

\section{Replica symmetry}
Within the replica symmetric approximation one assumes that the
solution of (\ref{eq:saddle}) is symmetric under permutations of the
replica indices implying that the saddle-point value of $c(\vs)$
depends only on the sum, $s:=\sum_a S^a$, of the components of the
vector $\vs$. It is then convenient to determine the distribution of
local magnetic fields $P(h)$ from its relation to $c(s)$ as given by
\cite{Remi} 
\begin{equation}
  \label{eq:Ph}
  c(s)=\int dh\; P(h)\; e^{-ikhs} \qquad\qquad 
  P(h)=\int \frac{ds}{2\pi}\; e^{ish}\; c(\frac s k) \; .
\end{equation}
Note that the $P(h)$ defined in this way is normalized only after the
limit $n\to 0$ is taken. The distribution of local magnetic fields is
equivalent to the 
free energy $f(\beta)$ since all thermodynamic properties may be
derived from suitable averages with $P(h)$ \cite{MePa2}. 

To get an equation for $P(h)$ from (\ref{eq:saddle})
we need to calculate 
\begin{align}
\sum_{\vs} e^{-ikhs}|\vs\cdot\vsi|^\al&=
\int\frac{dr\,d\rh}{2\pi} |r|^\al e^{ir\rh} 
\sum_{\vs}\exp\Big(-ikhs-i\rh\vs\cdot\vsi\Big)\\
&=\int\frac{dr\,d\rh}{2\pi} |r|^\al e^{ir\rh} 
\sum_{\vs}\prod_a \exp\Big(-iS^a(kh+\rh\sigma^a)\Big)\\
&=\int\frac{dr\,d\rh}{2\pi} |r|^\al e^{ir\rh} 
[2\cos(kh+\rh)]^{\frac{n+\sigma}{2}}\;[2\cos(kh-\rh)]^{\frac{n-\sigma}{2}}\\
&\rightarrow \int\frac{dr\,d\rh}{2\pi} |r|^\al e^{ir\rh} 
\left[\frac{\cos(kh+\rh)}{\cos(kh-\rh)}\right]^{\frac\sigma 2}\; ,
\end{align}
where the limit $n\to 0$ was performed in the last line and 
$\sigma:=\sum_a\sigma^a$. Using $\Lambda(n)\to 1$ for $n\to 0$
\cite{Remi} we therefore find from (\ref{eq:saddle}) in the replica
symmetric approximation 
\begin{equation}\label{eq:h1}
  c(\sigma)= \exp\left(-|k|^\al\int dh P(h)\int\frac{dr\,d\rh}{2\pi}
    |r|^\al \exp\Big(ir\rh +\frac\sigma
    2\ln\frac{\cos(kh+\rh)}{\cos(kh-\rh)}\Big)\right)\; .
\end{equation}
Using this result in (\ref{eq:Ph}) and performing the transformations
$r\mapsto r/k, \rh\mapsto \rh k$ we get 
\begin{equation}
  \label{eq:Phrs}
  P(h)=\int\frac{ds}{2\pi}\exp
       \left(ish-\int dh' P(h')\int\frac{dr\,d\rh}{2\pi} |r|^\al
       \exp\Big(ir\rh+\frac s{2k}
       \ln\frac{\cos(kh'+k\rh)}{\cos(kh'-k\rh)}\Big)\right)\; .
\end{equation}
We are now in the position to continue this result back to real values
of the temperature by simply setting $k=i\beta$. In this way we find
the following self-consistent equation for the replica symmetric field
distribution $P(h)$ of a Levy spin-glass at inverse temperature $\beta$
\begin{equation}
  \label{eq:resrs}
  P(h)=\int\frac{ds}{2\pi}\exp
    \left(ish-\int dh' P(h')\int\frac{dr\,d\rh}{2\pi} |r|^\al
    \exp\Big(ir\rh-i\frac s{2\beta}
    \ln\frac{\cosh\beta(h'+\rh)}{\cosh\beta(h'-\rh)}\Big)\right)\; . 
\end{equation}

\section{Spin glass transition}
\begin{figure}
  \begin{center}
    \includegraphics[width=0.6\textwidth]{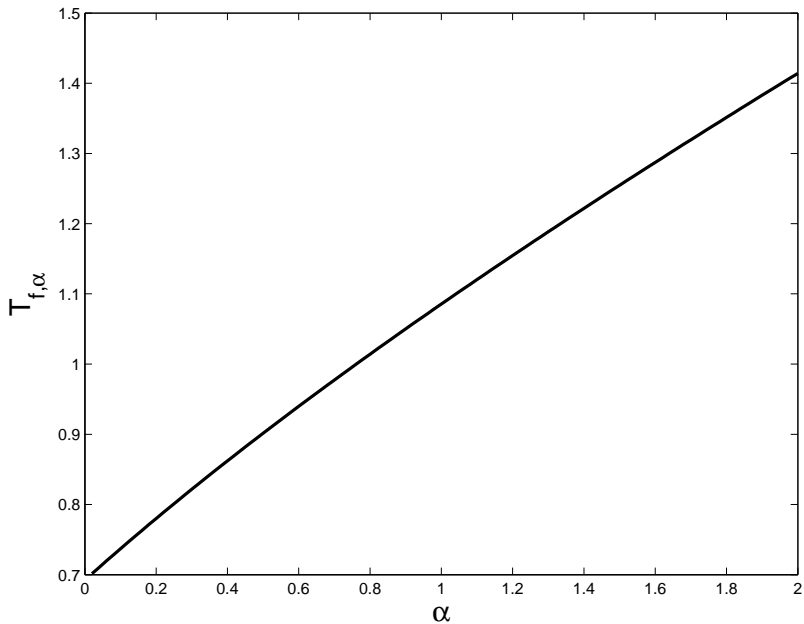}
  \end{center}
\caption{Freezing temperature $T_{c,\al}$ of an infinite-range
  spin-glass with Levy-distributed couplings as function of the
  parameter $\al$ of the Levy-distribution defined in
  (\ref{eq:defP}). For the scaling of the coupling strength with $N$
  as chosen in (\ref{eq:H}) there is a finite transition temperature
  for all values of $\al$. In the limit $\al\to 2$ the result for the
  SK-model is recovered.}
\label{f.1}
\end{figure}

From (\ref{eq:resrs}) we infer that the paramagnetic field
distribution,  $P(h)=\delta(h)$, is always a solution. To test its
stability we plug into the r.h.s. of (\ref{eq:resrs}) a distribution
$P_0(h)$ with a small second moment, 
$\epsilon_0:=\int dh P_0(h)\, h^2 \ll 1$, calculate the l.h.s. (to be
denoted by $P_1(h)$) by linearizing in $\epsilon_0$ and compare the new
second moment, $\epsilon_1:=\int dh P_1(h)\, h^2$, with
$\epsilon_0$. We find $\epsilon_1>\epsilon_0$, {\it i.e.} instability of the
paramagnetic state, if the temperature $T$ is smaller than a critical
value $T_{f,\al}$ determined by 
\begin{equation}
  (T_{f,\al})^\al=-\int\frac{dr\,d\rh}{2\pi} |r|^\al e^{ir\rh}
  \tanh^2 \rh
   =-\frac{\Gamma(\al+1)}{\pi}\,\cos(\frac{\al+1}{2} \pi)
      \int\frac{d\rh}{|\rh|^{\al+1}} \tanh^2 \rh .
\end{equation}
This result for the freezing temperature is essentially the same as
the one obtained by CB using the cavity method \cite{CiBo}. Our
somewhat more detailed prefactor ensures that the limit $\al\to 2$
correctly reproduces the value $T_f^{SK}=\sqrt{2}$ of the SK-model
\cite{SK}. The dependence of $T_{f,\al}$ on $\al$ is shown in
fig.~\ref{f.1}. 

The peculiarities of the spin-glass transition in the present system
are apparent from the similarity between (\ref{eq:resrs}) and 
analogous results for strongly diluted spin glasses and disordered
spin systems on random graphs \cite{ViBr,Remi,MePa2}. To make this
analogy more explicit we rewrite (\ref{eq:resrs}) in a form that
allows to perform the $s$-integration to obtain
\begin{align}\nonumber
  P(h) &=\int\frac{ds}{2\pi} e^{ish}  
     \sum_{d=0}^\infty \frac{(-1)^d}{d!} \int \prod_{i=1}^d 
 \Big(dh_i P(h_i)\frac{dr_i\,d\rh_i}{2\pi} |r_i|^\al e^{ir_i\rh_i}\Big)
    \exp\Big(-i\frac s{2\beta}\sum_{i=0}^d
    \ln\frac{\cosh\beta(h_i+\rh_i)}{\cosh\beta(h_i-\rh_i)}\Big)\\
  &=\sum_{d=0}^\infty \frac{(-1)^d}{d!} \int \prod_{i=1}^d 
  \Big(dh_i P(h_i)\frac{dr_i\,d\rh_i}{2\pi} |r_i|^\al e^{ir_i\rh_i}\Big)
  \;\delta\Big(h-\frac{1}{\beta}\sum_{i=0}^d 
          \tanh^{-1}(\tanh\beta h_i\tanh\beta\rh_i)\Big)\; . 
\end{align}
This form of the self-consistent equation is similar to those
derived within the cavity approach for systems with locally tree-like
topology \cite{ViBr,Remi,HaWe} and may also form a suitable starting
point for a numerical determination of $P(h)$ using a
population-dynamical algorithm \cite{MePa2}.

\section{Discussion}
Infinite-range spin-glasses with Levy-distributed couplings are
interesting examples of classical disordered systems. The broad
variations in coupling strengths brought about by the power-law tails
in the Levy-distribution violate the Lindeberg condition for the
application of the central limit theorem and give rise to non-Gaussian
cavity field distributions with diverging moments. We have shown that
it is nevertheless possible to derive the replica symmetric properties
of the system in a compact way by using the replica method as 
developed for the treatment of strongly diluted spin glasses and
optimization problems \cite{Remi} which focuses from the start on the
complete distribution of fields rather than on its moments. 

Due to the long tails in the distribution of coupling strengths Levy
spin-glasses  interpolate between systems with many,
i.e. $\mathcal{O}(N)$, weak couplings per spin as the
Sherrington-Kirkpatrick model and systems with few,
i.e. $\mathcal{O}(1)$, strong couplings per spin as the Viana-Bray
model. The majority of the 
$N-1$ random interactions coupled to each spin are very weak (of order
$N^{-1/\al}$). These weak couplings will influence only the very low
temperature behaviour which may be expected to be similar to
that of the SK-model. On the other hand the largest of $N$ random
numbers drawn independently from the distribution (\ref{eq:defP}) is
of order $N^{1/\al}$ \cite{BoGe} and hence every spin also shares a
fraction of strong bonds, $J_{ij}=\mathcal{O}(1)$, which are for
$|J_{ij}|>1/\beta$ practically frozen. With decreasing
temperature a growing backbone of frozen bonds builds up that
percolates at the transition temperature $T_{f,\al}$  \cite{CiBo}.  
The mechanism for the freezing transition is hence rather different
from that operating in the Sherrington-Kirkpatrick model and resembles
the one taking place in disordered spin systems on random
graphs with local tree-structure. 

\acknowledgments I would like to thank Daniel Grieser, R\'emi Monasson
and Martin Weigt for clarifying discussions.

\end{document}